\RequirePackage[fleqn]{amsmath}
\documentclass[12pt]{iopart}

\usepackage{url,graphicx,epstopdf,bm}
\usepackage{epsfig,hyperref,latexsym, dsfont, wasysym,multirow,mathrsfs,xcolor,color}
\usepackage[caption=false]{subfig}
\usepackage{braket}
\usepackage{lineno,hyperref}
\usepackage{python}
\modulolinenumbers[5]
\usepackage{soul}

\usepackage{todonotes}

\begin{document}
\journal{ \NJP}

\title{Quantum control for high-fidelity multi-qubit gates}

\author{Raymond J. Spiteri and Marina Schmidt}
\address{Department of Computer Science, University of Saskatchewan,
	Saskatoon, Saskatchewan S7N 5C9, Canada}
\author{Joydip Ghosh}
\address{Department of Physics, University of Wisconsin--
	Madison, Madison, Wisconsin 53706, United States of America}
\author{Ehsan Zahedinejad}
\address{Institute for Quantum Science and Technology, University of Calgary, Calgary, Alberta T2N~1N4, Canada}
\address{1QB Information Technologies (1QBit),
	Vancouver, British Columbia V6C 2B5, Canada}
\author{Barry C. Sanders}
\ead{sandersb@ucalgary.ca}
\address{Institute for Quantum Science and Technology, University of Calgary,
	Calgary, Alberta T2N~1N4, Canada}
\address{Program in Quantum Information Science, Canadian Institute for Advanced Research,
	Toronto, Ontario M5G 1M1, Canada}

\begin{abstract}
  Quantum control for error correction is critical for the practical
  use of quantum computers.  We address quantum optimal control for
  single-shot multi-qubit gates by framing it as a feasibility problem
  for the Hamiltonian model that is then solved with standard global
  optimization software.  Our approach yields faster high-fidelity
  ($>$99.99\%) single-shot three-qubit-gate control than obtained
  previously, and it has also enabled us to solve the quantum-control
  problem for a fast high-fidelity four-qubit gate.
\end{abstract}


\submitto{\NJP}

    
\section{Introduction}
Quantum computing could convert certain intractable computational
problems, such as number factorization, into efficiently solvable
problems by replacing binary representations and Boolean logic with
quantum-bit (qubit) strings and quantum gates~\cite{NC10}.  A
procedure in a quantum algorithm is expressed as a quantum circuit
comprising a universal primitive instruction set of quantum gates,
such as the Hadamard (H), $\pi/8$ (T), and two-qubit entangling
controlled-not (CNOT) gates, or, alternatively, the H gate along with
the three-qubit Toffoli gate, which effects controlled-controlled-not
(CCNOT) operation~\cite{NC10}.  Scalable quantum computing is achieved
if its components, such as preparation of qubits, quantum logic gates,
and measurements, achieve a minimum performance threshold, typically
expressed in terms of average gate fidelity but rigorously required to
be the diamond-norm-based error rate for the fault-tolerance
theorem~\cite{Got98,SWS16}.  If the performance threshold is met, then
the quantum information is encoded into a quantum codespace, processed
in that codespace by suitably encoded quantum gates, and decoded at
the output.

Although the Solovay--Kitaev theorem guarantees efficient
decomposition of quantum gates into a primitive instruction
set~\cite{DN06}, concatenating faulty gates unfortunately compounds
error. Fortunately, however, these errors can be ameliorated by
fault-tolerant methods~\cite{Got98}.  Even if high-quality gates are
created, the multi-qubit operations to map quantum information into
and out of the quantum codespace could fail due to accumulated errors.
Our aim is to devise versions of single-shot multi-qubit
gates~\cite{LBA+09,SFW+12,RDN+12,MKH+09}, where single-shot refers to
a controlled, uninterrupted, continuous-time evolution to realize the
quantum gate.  We solve for such gates based on a Hamiltonian
description of the physical system such that a suitable time-dependent
driving term delivers a high-fidelity approximation to the desired
quantum gate using error avoidance to achieve the requisite threshold
for efficient quantum error correction. Fast single-shot multi-qubit
gates are vital to achieving high fidelity because they are able to
operate faster than the decoherence time. The alternative to
decomposing a multi-qubit gate into a sequence of one- and two-qubit
gates is necessarily much slower and thus is affected significanlty by
decoherence.

To this end, some of us have designed an algorithm based on
differential evolution that delivers a fast high-fidelity Toffoli
gate~\cite{ZGS15} and other fast high-fidelity three-qubit
gates~\cite{ZGS16} in contrast to alternative approaches that
decompose a three-qubit gate into single- and two-qubit gates.  We use
average gate fidelity because the figure of merit for the quality of
the gate as this figure of merit is widely used and amenable to
experimental testing.  Whereas this approach is shown to deliver fast
high-fidelity three-qubit gates, extending to four-qubit gates was
infeasible for our available computational resources, and adoption of
this approach by other researchers was impeded by the need to learn
and code our specialized algorithm that we dubbed subspace-selective
self-adaptive differential evolution (SuSSADE)
algorithm~\cite{ZGS16,PWZ+17}.

Here we report a new method that is superior in that it reproduces
previous results, yields faster, higher-fidelity three-qubit gates,
and solves fast, high-fidelity four-qubit gates.  Our method is also
superior in that we use standard optimization software rather than
specialized in-house code, thus making our method easily accessible to
other researchers.  
Solving the quantum-control procedure in this way takes patience as
many runs are required and the runs can be slow, but our demonstration
of better three-qubit gates and the first solution for a single-shot
four-qubit gate (rather than decomposing the four-qubit gate into
concatenated fewer-qubit gates~\cite{Nigmatullin2009,MSG+11,FSB+12}) via
quantum control proves that this method works and is viable with
modest computational resources.

We have formulated the quantum-control problems as feasibility
problems that define the solution only in terms of criteria (i.e.,
design constraints) that the multi-qubit gate must possess.  These
criteria are typically expressed mathematically as sets of equations
or inequalities.  The feasibility problem is then solved using
standard global optimization software.  Most quantum-control problems
are solved using greedy algorithms~\cite{MSG+11}, which either rely on
the optimization problem being convex or on local solutions being good
enough, but our aim for
high-fidelity performance renders the response surface highly complex,
thus making global optimization algorithms
attractive~\cite{ZSS14}. Thus, our approach offers two advantages: the
first is using global optimization to obtain better pulse sequences
for gate optimization, and the second is avoiding the decomposition into
fewer-qubit gates that would slow operation and thus enhance the
deleterious effect of decoherence.

As an application of our quantum-controlled multi-qubit gate approach,
we focus on a superconducting-circuit realization because remarkable
progress has been achieved with superconducting artificial atoms,
realized as transmons or similar, such as the demonstration of nine
coherently coupled superconducting transmons~\cite{BSL+17}.  A
high-fidelity two-qubit entangling gate is achieved via exploiting
energy levels beyond the two-level qubit space and employing the
method of avoided level crossings~\cite{GGZ+13}.  The essence of the
avoided-level-crossing-level gate is that the frequency of each qubit
is tuned such that energy levels approach each other but remain
non-degenerate through the whole evolution.  Avoided-level-crossing
evolution mixes the energy population and dynamical phases such that
the final evolution of the system leads to the desired quantum gates.

Our strategy for designing quantum control to achieve multi-qubit
gates is based on avoided level crossings but for a higher
dimension~\cite{ZGS15,ZGS16}.  We first pose quantum control for
avoided level crossings as a feasibility problem.  Then we describe
how we use the Global Optimization Toolbox in
MATLAB\textsuperscript{\tiny\textregistered} to solve the feasibility
problem and extract the external pulses. We characterize the accuracy
of quantum gates by the intrinsic fidelity, from which the error rate
can be inferred.  A quantum gate design is defined as feasible if its
intrinsic fidelity satisfies a sufficiently high threshold, which here
is taken to be 99.99\%.

The outline of our paper is as follows. 
The superconducting-circuit model
for realizing a Toffoli gate is described in \S\ref{sec:model} by giving
a mathematical description of the mechanics involved.
Avoided-level-crossing-based quantum gates are described in detail in
\S\ref{sec:AvoidedCross}. This description is followed by an
explanation of the quantum-control procedure and the computational
methods involved in finding high-fidelity procedures in
\S\ref{sec:qControl}. Results are presented in
\S\ref{sec:results}
. Conclusions and directions for future
research are given in \S\ref{sec:conclusions}.

\section{Model}
\label{sec:model}
We consider~$n$ coupled superconducting artificial
atoms~\cite{BKM+14}, with parameters appropriate for capacitively
coupled transmon systems~\cite{KYG+07,HKD+09}.
Each transmon has $j$ energy states,
and transmon locations are denoted by $k\in[n]:=\{1,2,\dots,n\}$.
Anharmonicities of the second and third energy levels are
parametrized by~$\eta$ and~$\eta'$,
where we assume
\begin{equation}
	\eta=200~\text{MHz},\;
	\eta'\approx3\eta,
\end{equation}
which is appropriate under the cubic
approximation of the potential well~\cite{GGZ+13}.
Capacitive coupling between the transmons yields an~$XY$ interaction between
adjacent transmons (in the rotating frame) with a coupling strength of
$g=30$~MHz.

In the ideal gate, a pulse is sent into the transmon system to correct
for errors.
The shifted frequency for the pulse applied to
transmon~$k$ is the set of shifted frequencies
\begin{equation}
\label{eq:setshiftedfreq}
	\bm{\varepsilon}(t)
		=\{ \varepsilon_k(t) \}_{k=1}^n,\;
	k\in\{1,2,\dots,n\},\;
	t \in [0,\Theta],
\end{equation}
with each shifted frequency bounded by
\begin{equation*}
\label{eq:domain}
	-2.5~\text{GHz} \leq \varepsilon_k(t) \leq 2.5~\text{GHz}
\end{equation*}
in our numerical simulations.  The Hamiltonian for~$n$ capacitively
coupled transmons is represented as the $j^n$-dimensional
block-diagonal matrix~\cite{GGZ+13}
\begin{align}
\label{eq:qutritcontrol}
  \hspace*{23mm} \frac{{\bm{H}}(t)}{h} := \hat {\bm{H}}(t)
  &=\sum_{k=1}^{n} \mathcal{P}_{\bf I} ^{\otimes n}(
    \operatorname{diag} \left(
    0,\varepsilon_k(t),2\varepsilon_k(t)-\eta,
    3\varepsilon_k(t)-\eta'\right))\nonumber \\
  &\hspace*{10mm} +\frac{g}{2}\sum_{k=1}^{n-1}
    \mathcal{P}_{\bf I}^{\otimes (n-1)}\left(\bm{X}_k \otimes \bm{X}_{k+1}+\bm{Y}_k \otimes
    \bm{Y}_{k+1}\right),
\end{align}
where each block corresponds to a fixed number of excitations and acts
on the Hilbert space $\mathscr{H}_4^{\otimes n}$.  The promotion
operator $\mathcal{P}_{\bf A}^{\otimes n} ({\bf B})$ is defined as the
sum of all possible Kronecker products of {\bf B} with $(n-1)$ copies
of {\bf A}.

The coupling operators
\begin{equation}
	\bm{X}_k
		=\begin{pmatrix}0&1& 0&0\\1&0&\sqrt{2}&0\\0&\sqrt{2}&0&\sqrt{3}\\0&0&\sqrt{3}&0\end{pmatrix}_k, \;
      \frac{\bm{Y}_k}{\mathrm i}
  =\begin{pmatrix}0&-1& 0&0\\1&0&-\sqrt{2}&0\\0&\sqrt{2}&0&-\sqrt{3}\\0&0&\sqrt{3}&0\end{pmatrix}_k,
\end{equation}
are the generalized Pauli operators~\cite{GGZ+13}. Under the
assumption of uniform coupling, these operators are in fact
independent of~$k$.

The block-diagonal property of the Hamiltonian permits
Eq.~(\ref{eq:qutritcontrol}) to be reduced to a subspace in which at
most~$n$ excitations are present
\begin{equation}
\label{eq:Hp}
	\hat{\bm{H}}_p(t)
		=\mathscr{O}_n\hat{\bm{H}}(t)\mathscr{O}_n^{\dagger},
\end{equation}
for~$\mathscr{O}_n$ denoting the operator that truncates the
Hamiltonian~(\ref{eq:qutritcontrol}).
Hamiltonian~(\ref{eq:Hp}) evolution proceeds over the gate time~$\Theta$
such that the resultant unitary operator is
\begin{equation}
\label{eq:U}
\bm{U}\left(\Theta\right)
=\mathcal{T}\exp\left\{\int_0^\Theta\text{d}t\,
  \frac{{\bm{H}}_p(t)}{\text{i} \hbar} \right\}
=\mathcal{T}\exp\left\{-2\pi\text{i}\int_0^\Theta\text{d}t\,
  \hat{\bm{H}}_p(t)\right\},
\end{equation}
with~$\mathcal{T}$ the time-ordering operator~\cite{DGT86}. 

Transmon states are defined over a computational subspace~$\bm{U}\left(\Theta\right)$
of at most three excitations, where the resulting projection
\begin{equation}
\label{eq:pUp}
	\bm{U}_\text{cs}\left(\Theta\right)
		=\mathscr{P}\bm{U}\left(\Theta\right)\mathscr{P}^{\dagger},
\end{equation}
yields a computational subspace for unitary operator~\eqref{eq:U} with
(matrix) dimension~$2^n$.  The projected unitary
operator~\eqref{eq:pUp} is used to determine the intrinsic fidelity by
comparing it to the target gate $\bm{U}_\text{target}$ of the system.
However, for superconducting artificial atoms, local~$z$ rotations can
be performed quickly and accurately; therefore, we construct a
local~$z$-equivalence class for an arbitrary element (gate) in SU(4)
and then specialize to the ideal gate.

Subspaces~$\bm{U}\left(\Theta\right)$ and~$\bm{U}'\left(\Theta\right)$
of SU(4) are equivalent if
\begin{equation}
	\bm{U}'\left(\Theta\right)
		=\bm{U}_l\bm{U}\left(\Theta\right)\bm{U}_r,
\end{equation}
with $2^n$-dimensonal diagonal matrix
\begin{equation}
\label{eq:eqv}
	\bm{U}_{l,r}\left(\beta_1,\beta_2,\dots, \beta_k\right)
		:=\bigotimes_{\ell=1}^k\bm{R}_z(\beta_k),\;
	\bm{R}_z(\beta_k)
		=\operatorname{diag}
			\left(1,\text{e}^{-\text{i}\beta_k}\right)
\end{equation}
for~$\beta_k$ the angle between diagonal entries of unitary
evolution matrix~\eqref{eq:U}.
For example, the diagonals of $\bm{U}_{l,r}$
for the three-qubit case are represented explicitly as
\begin{align}
\label{eq:smatrix}
	\hspace*{12mm} \bm{U}_{l,r}
		=\text{diag}\big(1,\text{e}^{-\text{i}\beta_3},\text{e}^{-\text{i}\beta_2},
  \text{e}^{-\text{i}\left(\beta_2+\beta_3\right)},\text{e}^{-\text{i}\beta_1},
	\text{e}^{-\text{i}(\beta_1+\beta_3)},\text{e}^{-\text{i}(\beta_1+\beta_2)},\text{e}^{-\text{i}(\beta_1+\beta_2+\beta_3)}\big).
\end{align}
This matrix~(\ref{eq:smatrix}) is employed to perform phase compensation in our
numerical simulation.

By comparing the gates, the intrinsic fidelity is given by the trace formula
\begin{equation}
\label{eq:fidelity}
	\mathcal{F}(\Theta,n;\bm{\varepsilon}(t))
		=\frac{1}{2^n}\Big|\operatorname{tr}\big(\bm{U}_{\text{target}}^{\dagger}\bm{U}_\text{cs}\left(\Theta\right)\big) \Big|\in[0,1]
\end{equation}
and is a
performance figure of merit for the quantum-gate operation for given
shifted frequencies that does not include decoherence.
Our aim is to find a set of shifted frequencies~(\ref{eq:setshiftedfreq})
such that
\begin{equation}
	\mathcal{F}(\Theta,n;\bm{\varepsilon}(t)) \ge 99.99\%
\end{equation}
for
multi-qubit gates of size $n=3,4$.
The set~(\ref{eq:setshiftedfreq}) of shifted frequencies
is defined to be the solution to a feasibility problem.

\section{Avoided-level-crossing-based quantum gates }
\label{sec:AvoidedCross}
We now explain the avoided-level-crossing-based Controlled-Z
(CZ) gate for the
physical model of two capacitively coupled frequency-tunable
transmons, with~$\eta$ and~$g$ ($g \ll \eta$) being the anharmonicity
and coupling strength, respectively. We clarify why finding a
theoretical solution for multi-qubit gates is challenging and motivates our quantum-control approach.

Whether in the sudden or the adiabatic regime, the idea of engineering
a pulse for the avoided-level-crossing-based gate in a superconducting
architecture remains the same. To design a control pulse for the two-qubit 
CZ gate, we tune the transmon frequencies such that the
$\ket{11}$ state mixes with the other states in the second excitation
subspace while the states in zero- and single-excitation subspaces
remain detuned from each other. However, for practical
implementations, both the sudden and adiabatic regimes are unsuitable
for obtaining the threshold fidelity required for the fault tolerance
(at least $99.99\%$ for topological error correction~\cite{GFG12}), and one needs
to resort to advanced quantum-control techniques to engineer the
feasible pulse; this is the primary motivation for our work.

One idea for designing three-qubit gates is to couple three transmons
via a superconducting cavity, usually referred to as the
circuit-quantum-electrodynamics (cQED) architecture~\cite{WSB+04} and
tune the transmon frequencies in the dispersive regime such that the
time-evolution operator gives rise to the target three-qubit gate at the
end of the operation.
Whereas such an approach has already been used to
demonstrate a Toffoli gate~\cite{RDN+12}, we do not consider cQED
hardware here because the architecture only contains a few transmons
inside a superconducting cavity and is thus not evidently scalable.

Instead, we consider a one-dimensional chain of~$n$
  transmons with nearest-neighbour coupling.
 One might seek to avoid the advanced quantum control
  approach and resort to a theoretical approach for designing 
  multi-qubit gates~\cite{GGZ+13},
  but this approach could fail for designing multi-qubit gates.
In our approach,
we initially detune the transmon frequencies from each other, which
  makes all eigenstates of the system non-degenerate.
  In order to
  obtain a high-fidelity three-qubit gate, we then vary the
  frequencies of all the qubits simultaneously such that the
  $\ket{111}$ state mixes strongly with other states in the third
  excitation subspace, while all other computational-basis states
  are detuned from each other.

\begin{figure}[htbp]
\centering
	\includegraphics[width=0.7\textwidth]{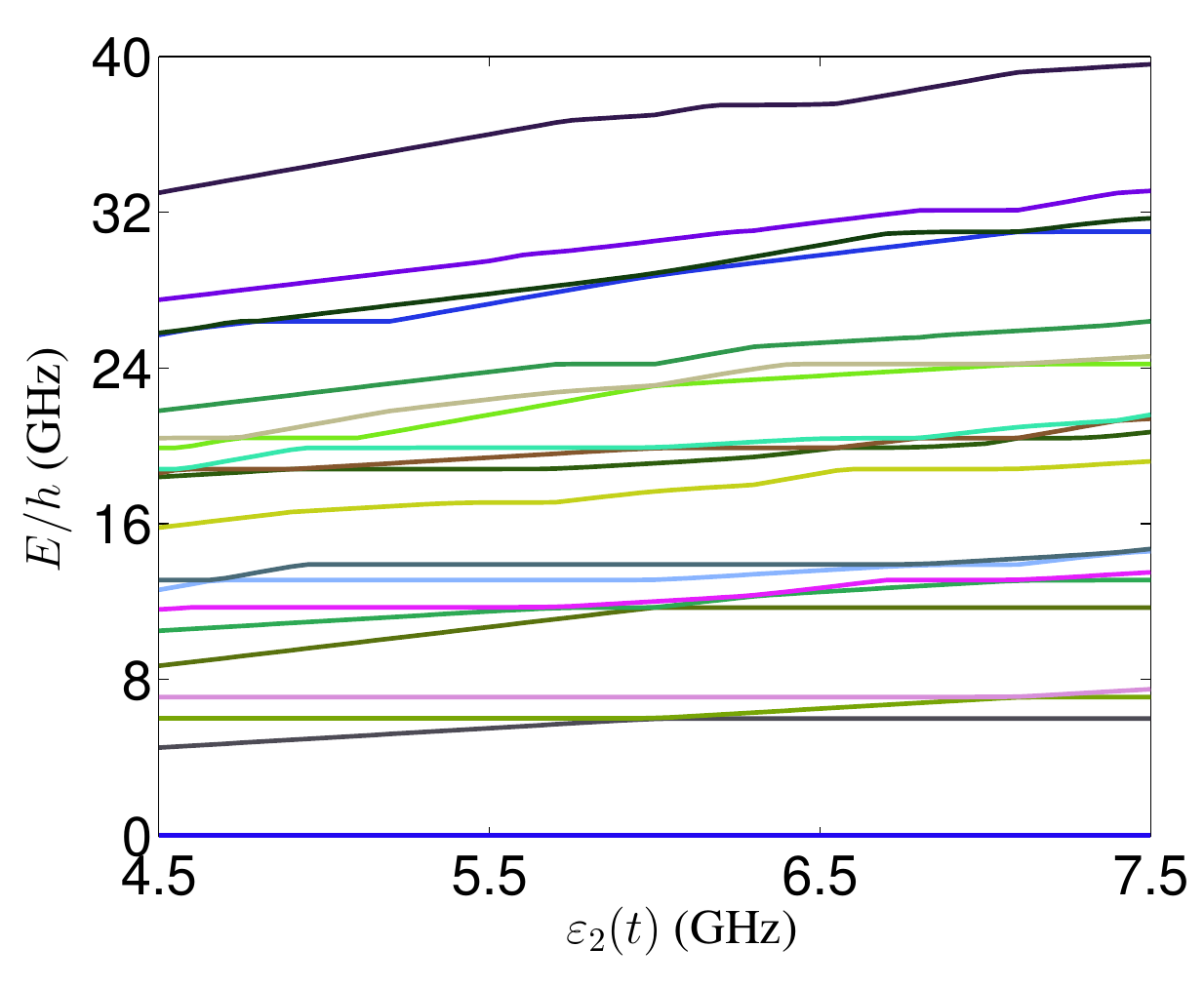}
        \caption{Energy spectrum of three
          nearest-neighbour-coupled transmons, where energy levels that have at most 3 excitations are shown. The first and third
          transmon frequencies are fixed at $4.8$~GHz and $6.8$~GHz,
          respectively. The frequency of the second transmon varies
          from $4.5$~GHz to $7.5$~GHz.}
\label{fig:Avoided_C_CZ}
\end{figure}

  Fig.~\ref{fig:Avoided_C_CZ} shows the energy spectrum of the
  three-transmon Hamiltonian (truncated up to the third excitation
  subspace) as a function of the frequency of the second transmon
  while other frequencies are kept fixed at 4.8 GHz and 6.8 GHz.  It
  is clear from Fig.~\ref{fig:Avoided_C_CZ} that devising an analytic
  quantum-control technique for a three-transmon system where all the
  transmons are driven simultaneously is a difficult task because of
  the existence of so many avoided crossings in the energy spectra.

We can follow the same argument to explain an
avoided-level-crossing-based approach for designing a four-qubit
gate. For this case, we initially detune the frequency of four
individual transmons to certain finite values and then drive all of
them simultaneously such that the $\ket{1111}$ state mixes strongly
with the other states in the fourth excitation subspace, with all
other computational-basis states detuned from each other.

From what we have described for a three-qubit gate, the
avoided-level-crossing-based four-qubit-gate design problem is
evidently much harder.  For the case of the three-qubit gate, the
truncated Hamiltonian includes the first 20 levels of the energy
spectrum, whereas this number is 66 for a four-qubit gate. The
increased number of energy levels increases the dimension of the
optimization problem for the design of the four-qubit gate, and the
complexity in solving optimization problems scales exponentially with
dimension. For example, the increased number of excitations in each
subspace also makes it significantly more difficult relative to two-
or three-qubit gates to find an optimal region to establish a working
avoided-level-crossing-based procedure to design four-qubit
gates. This is why we employ a quantum-control scheme to devise a
policy for designing such multi-qubit gates~\cite{ZGS15}.

\section{Quantum control}
\label{sec:qControl}

The following subsections discuss how our quantum-control procedure is
realized. Approaches in the literature have formulated the problem as
an optimization problem.
The procedure here is based on finding a solution to a feasibility
problem for an~$n$-qubit circuit where shifted pulse frequencies are
chosen such that the only requirement is that the intrinsic fidelity
meets or exceeds a given threshold. We describe how the feasibility
problem is stated and solved using the GlobalSearch solver in in
MATLAB\textsuperscript{\tiny\textregistered} for the three-qubit and
four-qubit cases.

\subsection{Feasibility problem}
We formally state the feasibility problem for the design of a
high-fidelity~$n$-qubit gate as follows.
Given a circuit with~$n$
qubits and a gate time~$\Theta$,
\begin{subequations}
\label{eq:feasibility}
\begin{align}
  \text{find} & & & \ {\bm \varepsilon}(t),  & t &\in [0,\Theta], \\
\label{eq:domain}
  \hspace*{10mm} \text{subject to} & & -2.5\ \text{GHz} \leq & \ \varepsilon_k(t) \leq 2.5\
                      \text{GHz}, & k&=1,2,\ldots,n, \\
\label{eq:constraint} 
             & & \mathcal{F}\left(\Theta,n;\bm \varepsilon(t)\right) & \geq 99.99\%.
\end{align}
\end{subequations}

The response surface defined by the average gate
fidelity~\eqref{eq:fidelity} is highly non-convex.  Despite this
issue, approaches based on local (greedy) algorithms such as GRAPE and
CRaB are mentioned as preferred approaches to solving quantum control
problems~\cite{EQTR2017}.  Local algorithms such as
Nelder--Mead~\cite{NelderMead1965}, Krotov~\cite{Krotov1995}, and
quasi-Newton~\cite{Fle13} fail~\cite{ZSS14}. Accordingly, we turn to
global optimization strategies to search the domain~\eqref{eq:domain}
of the shifted frequencies to obtain a feasible solution.

The basic globalization strategy of employing a multiple-restart (or
multi-start) of a (local) quasi-Newton optimization has
advantages~\cite{Langbein2015} with respect to quality of solution and
required computational resources over greedy methods.  However,
multi-start methods and genetic algorithms were ultimately ineffective
in finding solutions to~\eqref{eq:feasibility}~\cite{ZSS14}.  The most
effective approach~\cite{ZSS14} is based on differential
evolution~\cite{Storn1997} and led to an effective approach dubbed
Subspace-Selective Self-Adaptive DE (SuSSADE))~\cite{ZGS15}.

Discrepancies between algorithmic performance is frequent in
scientific computing; robustness and speed often must be traded off
for each other, where robustness in this context refers to the
reliability of obtaining a consistent solution. On the one hand,
purely global optimization algorithms such as SuSSADE are slow and
steady. In the long run, they are expected to give the best results as
defined by the final value of the objective function. On the other
hand, local-optimization strategies such as multi-start can be
relatively fast and give good answers.  The final values of the
objective function, however, are typically inferior to solutions
provided by purely global optimization algorithms applied to
non-convex optimization problems, given sufficient time and
computational resources.

The frequency components~(\ref{eq:setshiftedfreq})
in~\eqref{eq:feasibility} are represented as piecewise-constant (step)
functions with time step duration $\Delta t=1$ ns.  The gate time
$\Theta$ is assumed to be an integer number of $\Delta t$ for a given
simulation.  Thus, each feasibility problem~(\ref{eq:feasibility}) has
associated with it $n\Theta/\Delta t$ degrees of freedom from
(\ref{eq:domain}) that are used to satisfy
Inequality~(\ref{eq:constraint}). The number of degrees of freedom is
the same as the dimension of the control space. It is a
straightforward observation that decreasing the time step duration
$\Delta t$ or increasing the gate time~$\Theta$ results in more
available degrees of freedom and hence a greater chance of finding a
solution to~(\ref{eq:feasibility}) for a given number of
transmons~$n$. However, there are limits on how small $\Delta t$ can
be as well as how large~$\Theta$ can be for viability in
practice. This, combined with the fact that the degree of
non-convexity of the response surface increases dramatically with~$n$,
makes~\eqref{eq:feasibility} a challenging problem to solve.

In principle, feasibility problems do not have a formal objective
function, in that sense distinguishing them from optimization
problems. In practice, however, feasibility problems are typically
solved as optimization problems. Often, an objective function is
created as a penalty function of constraint violations, e.g., the sum
of the squares of the constraint violations. The idea is that in the
face of multiple (properly scaled) constraints, the solution is not
biased by an objective function when trying to find a solution to a
feasibility problem. For a given problem, it may not make sense to
optimize anything, and in fact doing so may hinder or even prevent
finding a feasible solution.
In the special case of~\eqref{eq:feasibility}, 
with only one constraint
(\ref{eq:constraint}),
efficiency of the
search can in practice be enhanced
through minimizing an appropriate
objective function such as infidelity
\begin{equation*}
	1-\mathcal{F}\left(\Theta, n; \bm{\varepsilon}(t)\right)
\end{equation*}
 or the angular deviation in Hilbert space from the target gate
\begin{equation}
\label{eq:arccos}
	\arccos(\mathcal{F}(\Theta, n; \bm \varepsilon(t))).
\end{equation}
We find angular deviation~\eqref{eq:arccos}
to yield the fastest convergence. 

\subsection{Algorithm}
Here, the three-qubit and four-qubit cases are optimized using the
GlobalSearch solver from the Global Optimization Toolbox in
MATLAB\textsuperscript{\tiny\textregistered}.  The optimization is
subject to a non-linear constraint arising from the feasibility
condition~\eqref{eq:constraint}. This solver is based on the scatter
search / nonlinear programming solver as implemented
in~\cite{Ugray2007}.  Scatter search can be thought of as a
sophisticated multi-start algorithm for solving global optimization
problems; i.e., the restart points are chosen adaptively as the
information about the response surface becomes available. In contrast,
classical multi-start algorithms, such as that described
in~\cite{Langbein2015} or the MultiStart algorithm in
MATLAB\textsuperscript{\tiny\textregistered}, simply start local
optimization algorithms from uniformly distributed or random points in
the design space.

Although in principle a sensible global optimization procedure should
eventually converge to a global optimum, the amount of time and
computational resources required to realize such convergence may not
be realistic.  Besides using GlobalSearch, we experimented with a
number of global optimization solvers: SuSSADE, the DIRECT
algorithm~\cite{Jones1993}, particle swarm
optimization~\cite{Kennedy1995} (and variants thereof), and
MultiStart.  We found GlobalSearch to be the most effective by a
rather wide margin: it was the only algorithm that was ultimately able
to find previously undiscovered feasible solutions. Optimization
instances for three-, four-, or five-qubit circuits with
$\Theta \le 100$~ns typically require only about 500 MB of RAM,
meaning that from this perspective they can be run using modest
hardware, such as a laptop computer. Serial computation times required
to find solutions vary considerably, from a few minutes for the
solution of the three-qubit problem with $\Theta = 26$~ns to a few
weeks for the four-qubit problem with $\Theta = 70$~ns to months for
the three-qubit problem with $\Theta = 24$~ns. Parallel processing
would speed up these times, the extent to which depends on the number
of processors available. In such situations, high-performance
computing hardware, e.g., in the form of clusters, would be more
appropriate than desktop computing.

Beyond the goal of solving the feasibility problem defined
by~(\ref{eq:constraint}), we are also interested in devising gates
with short gate times~$\Theta$. Note that, although we seek small
values of~$\Theta$, our approach is not optimization per se but rather
employs a constraint on fidelity for fixed reasonable gate time to
compute a solution.  Posing the problem as finding the shortest gate
time subject to the constraint of sufficiently high fidelity is
equivalent to a single-objective optimization problem. In our work,
the gate time is reduced directly by attempting to solve the
feasibility problem for smaller and smaller~$\Theta$ values until
solutions could no longer be found.  A form of parameter continuation
was employed such that feasible solutions for longer gate times were
used as starting points in the solver for shorter gate times in the
global search algorithm. This enabled more robust and more efficient
convergence to a solution.

\section{Results}
\label{sec:results}
\begin{figure}[htbp]
\centering
\includegraphics[width=0.7\textwidth]{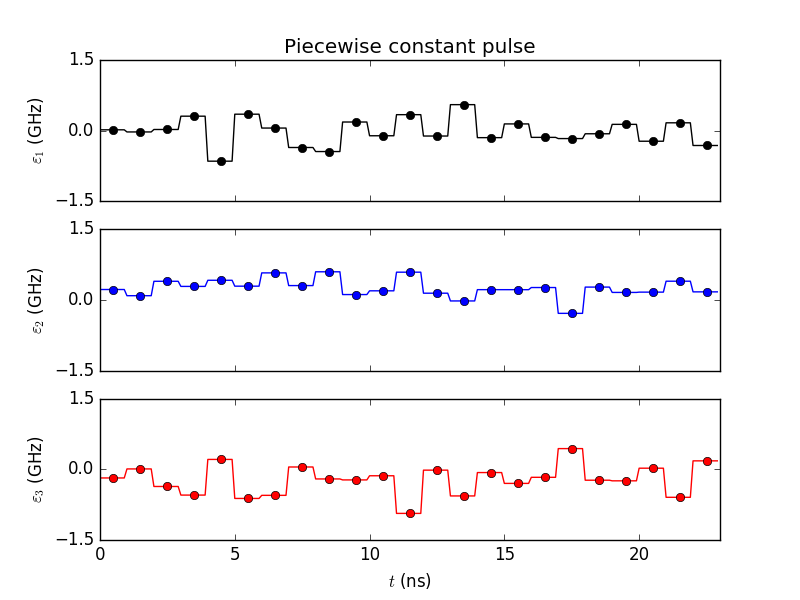}
\caption{Piecewise-constant pulse representation of the frequencies
  $\varepsilon_{k\in\{1,2,3\}}$ for the superconducting atoms versus
  the evolution time $t$. The size of time step is
  $\Delta{t}=1\ \text{ns}$. 
  The generated pulses enable the design of a high-fidelity
  ($\mathcal{F}(\Theta,n;\bm{\varepsilon}(t))\ge99.99\%$) Toffoli gate
  that operates in $23\ \text{ns}$. The solid dots show the control
  parameters used through the feasibility process to tune the shape of
  the pulses.}
\label{fig:three_qubit_square}
\end{figure}

Using the quantum-control process described in
\S\ref{sec:qControl}, pulses are generated for the design of a single-shot
high-fidelity three-qubit Toffoli gate that lead to a
feasible solution of~\eqref{eq:feasibility} over the minimal duration
time of $23\ \text{ns}$. This is an improvement of 11.5\% over the
gate time of $26\ \text{ns}$ reported
in~\cite{ZGS15}. Figure~\ref{fig:three_qubit_square} shows the
resulting piecewise-constant pulses as a function of $t$.

\begin{figure}[htbp]
\centering
\includegraphics[width=0.7\textwidth]{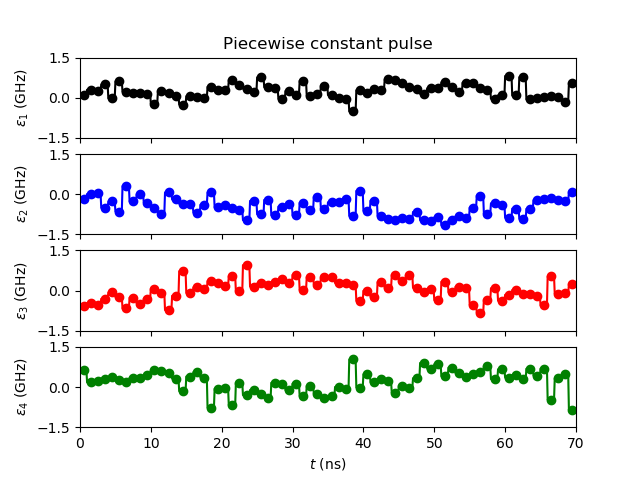}
\caption{Piecewise-constant pulse representation of the frequencies
  $\varepsilon_{k\in\{1,2,3,4\}}$ for the superconducting atoms versus
  the evolution time $t$. The size of time step is
  $\Delta{t}=1\ \text{ns}$. 
  The generated pulses enable the design of a high-fidelity
  ($\mathcal{F}(\Theta,n;\bm{\varepsilon}(t))\ge99.99\%$) CCCZ gate
  that operates in $70\ \text{ns}$. The solid dots show the control
  parameters used through the feasibility process to tune the shape of
  the pulses.}
\label{fig:four_qubit_square}
\end{figure}

The quantum-control process is also able to produce a solution to the
feasibility problem~\eqref{eq:feasibility} for a four-qubit CCCZ
gate. Figure~\ref{fig:four_qubit_square} shows the pulses that lead to
a feasible solution of~\eqref{eq:feasibility} for
piecewise-constant 
pulses as a function of $t$.  
The resultant four-qubit gate operates in $70\ \text{ns}$. 
This is the first design to satisfy~\eqref{eq:feasibility} with $n=4$.
Suitably designed single-shot four-qubit gates could be valuable for
four-qubit encoding to correct one erasure error~\cite{GBP97} and sets
the stage for progressing to higher-order single-shot multiqubit gates
such as a five-qubit gate that could encode in a single shot for the
five-qubit code~\cite{LMPZ96}

\section{Conclusions}
\label{sec:conclusions}
We have formulated the problem of designing multi-qubit gates as a
feasibility problem. This allows for a formally proper problem
formulation because at its core the only defining feature of a
solution is that the gate has a sufficiently high intrinsic fidelity
($\mathcal{F}(\Theta, n; \bm{\varepsilon}(t)) \ge 99.99\%$). Because
feasibility problems are often solved as optimization problems and
there is only one such defining feature, convergence to a solution
to~\eqref{eq:feasibility} can be enhanced through the use of an
appropriate objective function with a non-zero gradient at the
solution. The control pulses for the transmon-shifted frequencies are
discretized using piecewise-constant representations in time steps of
1 ns. The degrees of freedom of the feasibility problem are the
parameters that represent the piecewise-constant values. Determining
small gate times~$\Theta$ is done directly by solving a feasibility
problem for fixed~$\Theta$ then reducing it until a feasible solution
could no longer be found. Extensive tests were performed for the
three-qubit system, so we have a high degree of confidence in the
minimality of the gating time attained. The testing with the
four-qubit system has thus far been more limited, and the problem is
much more challenging; hence we make no strong claims as to the
minimality of~$\Theta$ in this case.

We have employed this approach to design single-shot high-fidelity
quantum gates, including the Toffoli and CCCZ gates. The operation
time of these three- and four-qubit gates is comparable with the time
of entangling two-qubit CZ~\cite{BKM+14} gates under the same
experimental constraints. We have designed the multi-qubit gates for a
simple architecture of linearly capacitively coupled superconducting
atoms that can be a module of any 1D or 2D architecture.  Our approach
can be used in any quantum-control approach when the underlying
problem can be formulated as a feasibility problem.  Here our approach
assumes that Hamiltonian evolution provides a sound description of the
quantum dynamics. Noise and parameter variability may be important in
practice, however, in which case quantum master equations or other
descriptions could be used if the noise is fully
understood~\cite{ZGS15,ZGS16}; otherwise statistical or black-box
techniques such as reinforcement learning could be
adopted~\cite{PWZ+17}.

\section*{Acknowledgments}
RJS and MS gratefully acknowledge support from the Natural Sciences
and Engineering Research Council of Canada (NSERC) under RGPIN
228090-2013. JG acknowledges support from the Vannevar Bush Faculty
Fellowship program sponsored by the Basic Research Office of the
Assistant Secretary of Defense for Research and Engineering and funded
by the Office of Naval Research through Grant
No. N00014-15-1-0029. This work was also supported in part by Army
Research Office Grant No. W911NF-12-0607 and National Science
Foundation (NSF) Grant No. PHY-1104660.  BCS appreciates support from
AITF and NSERC.
\section*{References}
\bibliography{qControl}
\end{document}